\documentstyle[12pt,epsf]{article}
\newcommand{\be}{\begin{equation}}
\newcommand{\ee}{\end{equation}}
\newcommand{\bea}{\begin{eqnarray}}
\newcommand{\eea}{\end{eqnarray}}
\newcommand{\bdm}{\begin{displaymath}}
\newcommand{\edm}{\end{displaymath}}
\newcommand{\no}{\nonumber \\}

\newcommand{\ubar}{\overline{\rule[0.42em]{0.4em}{0em}}\hspace{-0.5em}u}
\newcommand{\dbar}{\,\overline{\rule[0.7em]{0.4em}{0em}}\hspace{-0.6em}d}
\newcommand{\sbar}{\overline{\rule[0.45em]{0.4em}{0em}}\hspace{-0.5em}s}

\newcommand{\qbar}{\overline{\rule[0.42em]{0.4em}{0em}}\hspace{-0.5em}q}

\newcommand{\eff}{{e\hspace{-0.1em}f\hspace{-0.18em}f}}

\newcommand{\sA}{s_{\hspace{-0.15em}A}}
\newcommand{\DGMO}{\Delta_{\mbox{\tiny GMO}}}
\newcommand{\mixingangle}{\theta_{\eta^\prime\hspace{-0.05em}\eta}}
\newcommand{\R}{{\scriptscriptstyle R}}
\renewcommand{\L}{{\scriptscriptstyle L}}
\newcommand{\al}{&\!\!\!\!}
\newcommand{\fs}{\; \; .}
\newcommand{\co}{\; \; ,}

\begin{document}

\begin{titlepage}

\begin{flushright}
BUTP-96/5\\
\end{flushright}

\begin{center}
{\LARGE {\bf \rule{0em}{4em}Implications of $\eta\eta^\prime$ mixing
for the decay $\eta\rightarrow3\pi$\rule{0em}{1em} }}\\ \vspace{0.8cm}
H. Leutwyler\\Institut f\"{u}r theoretische Physik der Universit\"{a}t
Bern\\Sidlerstr. 5, CH-3012 Bern, Switzerland and\\
CERN, CH-1211 Geneva, Switzerland\\
\vspace{0.6cm}
January 1996\\
\vspace{0.6cm}
{\bf Abstract} \\
\vspace{1.2em}
\parbox{30em}{Taken by itself, the interference with the
$\eta^\prime$ appears to strongly affect the amplitude of the
transition $\eta\!\rightarrow\!3\pi$. I point out that this
effect is fictitious and also occurs in the mass spectrum of the pseudoscalars.
Chiral symmetry implies
that the same combination of effective coupling constants which determines the
small deviations from the Gell-Mann-Okubo formula also specifies the symmetry
breaking effects in the decay amplitude and thus ensures that these are small.}
\\ \vspace{1em}
\rule{30em}{.02em}\\
{\footnotesize Work
supported in part by Schweizerischer Nationalfonds}
\end{center}
\end{titlepage}

The decay $\eta\rightarrow 3\pi$ is of particular interest, because it violates
isospin symmetry. The electromagnetic interaction is known to
produce only very small corrections \cite{Sutherland}. Disregarding these, the
transition amplitude is proportional
to $m_d-m_u$ and thus represents a sensitive probe of the
symmetry breaking generated by the quark masses.

Since the quark mass term $\qbar m q\!=\!m_u\,\ubar u+m_d\, \dbar d+m_s\,
\sbar s$ breaks SU(3), it generates transitions between the octet and the
singlet of pseudoscalar mesons. The consequences for the transition
amplitude are discussed in the literature, but the
results are contradictory: While a direct evaluation of the mixing
effects \cite{Akhoury Leurer Pich}
leads to the conclusion that the current algebra prediction is modified
drastically,
the chiral perturbation theory calculation to one loop \cite{GL
eta} yields the opposite result. The purpose of the present paper is to resolve
this paradox.

The direct calculation is based on the effective Lagrangian which describes the
low energy structure of QCD in terms of a simultaneous expansion in powers of
$1/N_c$, powers of the momenta $p$ and powers of the quark mass matrix $m$.
For a discussion of this framework to first nonleading order and
references to the literature, see ref.\cite{HL bound}.
The relevant effective field
$U(x)$ is an element of $\mbox{U(3)}$ and includes the
degrees of freedom of both the pseudoscalar octet and the singlet. The latter
is described by the phase of the determinant, $\mbox{det}U\!=\!e^{i\phi_0}$.
Counting the three expansion parameters as
small quantities of order
$1/N_c\!=\!O(\delta)\,,\;p\!=\!O(\sqrt{\delta})\,,\;
m\!=\!O(\delta)$, the
expansion starts with a contribution of order one, given by
\be\label{Leff}{\cal L}_\eff=
\mbox{$\frac{1}{4}$}F^2\langle\partial_\mu U^\dagger\partial^\mu
U\rangle +\mbox{$\frac{1}{2}$}F^2B\langle
m U^\dagger+m U\rangle-\mbox{$\frac{1}{2}$}\tau\phi_0^2+
O(\delta)\co\ee
where matrix traces are abbreviated with the symbol $\langle\ldots\rangle$.
The expression involves three effective coupling constants: the pion decay
constant $F\!=\!O(\sqrt{N_c})$, the constant $B\!=\!O(1)$ which
determines the
magnitude of the quark condensate and the topological susceptibility
$\tau\!=\!O(1)$. The relative magnitude of the three leading contributions
depends on the relative magnitude of the expansion parameters:
the first involves two powers of momentum, the second is proportional to the
quark mass matrix and the coefficient $\tau$ of the third is smaller than
$F^2$ or $F^2B$ by one power of $1/N_c$.

Setting $U\!=\!\exp i\varphi/F $, the singlet field is given by the trace
$\phi_0\!=\! \langle\varphi\rangle/F$, so that the
terms quadratic in $\varphi$ are $\frac{1}{4}\langle\partial_\mu
\varphi\partial^\mu\varphi\rangle-\frac{1}{2} B \langle m\varphi^2\rangle
-\frac{1}{2}\tau\langle\varphi\rangle^2/F^2$.
For those fields which carry
electric charge or strangeness, this expression is diagonal and leads
to the standard current algebra mass formulae,
\bdm M_{\pi^+}^2\!=\!(m_u+m_d)
B\co\;\;M_{K^+}^2\!=\!(m_u+m_s)B\co\;\;M_{K^0}^2\!=\!(m_d+m_s)B\fs\edm
The states
$\pi^0,\eta$ and $\eta^\prime$ undergo mixing. The mixing matrix is an element
of O(3) and may thus be represented in terms of three angles,
$\theta_{\eta\eta^\prime},\epsilon,\epsilon^\prime$. The first arises
from the mass difference between the strange and nonstrange quarks
and breaks
SU(3), while the other two are isospin breaking effects, driven by $m_d-m_u$.
To first order in isospin breaking, the relation between the neutral components
of the field,
$\varphi=\varphi_3\lambda_3+\varphi_8\lambda_8+\varphi_9\sqrt{\frac{2}{3}}$ and
the mass eigenstates $\pi^0,\eta,\eta^\prime$ is of the form
\bea \label{change of basis}
\varphi_3\al=\al\;\pi^0-\epsilon\;\eta-\epsilon^\prime\; \eta^\prime\\
\varphi_8\al=\al \;\;\;\cos \theta_{\eta\eta^\prime}\,(\eta+\epsilon\,\pi^0)
+  \sin \theta_{\eta\eta^\prime}\,(\eta^\prime+\epsilon^\prime\,\pi^0)\no
\varphi_9\al=\al -\sin \theta_{\eta\eta^\prime}\,(\eta+\epsilon\,\pi^0)
   +   \cos \theta_{\eta\eta^\prime}\,(\eta^\prime+\epsilon^\prime\,\pi^0)
\fs\nonumber\eea
In the eigenvalues, the isospin breaking effects are of order $(m_d-m_u)^2$.
Neglecting these, the $\pi^0$ is degenerate with $\pi^{\pm}$,
$M_{\pi^0}^2=(m_u+m_d) B$. The remaining two eigenvalues
involve a new
scale, set by the topological susceptibility. Eliminating it, the
diagonalization leads to two independent relations among the three quantities
$M_\eta,M_{\eta'},\theta_{\eta\eta'}$, e.g.
\bea\label{mixing a} &&\sin 2\theta_{\eta\eta'}
=-\mbox{$\frac{4}{3}$}\sqrt{2}\,\frac{M_K^2-M_\pi^2}
{M_{\eta^\prime}^2-M_\eta^2}\co\\
\label{mixing b}&& M_\eta^2=\mbox{$\frac{1}{3}$}(4M_K^2-M_\pi^2)+
\mbox{$\frac{2}{3}$}\sqrt{2}\;\mbox{tg}\,\theta_{\eta\eta'}\,(M_K^2-M_\pi^2)\fs
\eea
The isospin breaking angles $\epsilon,\epsilon'$ are proportional to
the quark mass ratio
\bdm \epsilon_0\equiv
\frac{\sqrt{3}}{\,4}\,\frac{(m_d-m_u)}{(m_s-\hat{m})}\co\;\;\;\;\;
\hat{m}=\mbox{$\frac{1}{2}$}(m_u+m_d)\fs\edm
The coefficients of proportionality may be expressed in terms of
$\theta_{\eta\eta'}$,
\bea\label{epsilon} \epsilon\al=\al\epsilon_0\,\cos\,
\theta_{\eta\eta'}\,\frac{\cos\,\theta_{\eta\eta'}-
\sin\, \theta_{\eta\eta'}\,\sqrt{2}}{\cos\,\theta_{\eta\eta'}+
\sin\, \theta_{\eta\eta'}/\sqrt{2}}\\
 \epsilon'\al=\al-2\,\epsilon_0\,\sin\,\theta_{\eta\eta'}\,
\frac{\cos\,\theta_{\eta\eta'}+
\sin\, \theta_{\eta\eta'}/\sqrt{2}}{\cos\,\theta_{\eta\eta'}-
\sin\, \theta_{\eta\eta'}\,\sqrt{2}}\fs\nonumber\eea
Note that, in the counting of powers introduced above,
$M_\pi^2,M_K^2,M_\eta^2,M_{\eta'}^2$
are treated as small quantities of order $\delta$. According to
eq.(\ref{mixing a}), the mixing angle $\mixingangle$ is given by a ratio
thereof and thus represents a quantity of order one. In this sense, the above
formulae are valid to all orders in $\mixingangle$. Numerically, the ratio
$M_{\eta^\prime}^2/M_{\eta}^2$ is about equal to 3, indicating that the
breaking of U(3)$_\R\times$U(3)$_\L$ generated by the anomaly is
larger than the breaking due to $m_s$, by roughly this factor. The
topological susceptibility, which describes the effects of the
anomaly in the framework of the effective Lagrangian and is of order
$N_c^{\,0}$, is more important than the terms of order $ m N_c$, which
account for the symmetry breaking generated by the quark masses.

Consider now the decay $\eta\rightarrow\pi^+\pi^-\pi^0 $. To calculate the
corresponding
transition amplitude with the effective Lagrangian in
eq.(\ref{Leff}), the expansion in powers of the field $\varphi$
is needed to order $\varphi^4$. The first term
yields a contribution proportional to
$\langle[\partial_\mu\varphi,\varphi][\partial^\mu \varphi,\varphi]\rangle$.
Since the remainder does not involve derivatives, the decay
amplitude $A$ involves at most two powers of momentum. Lorentz
invariance and crossing symmetry then imply that $A$ is of the form $a +
b\, s$, where $s\!=\!(p_{\pi^+}+p_{\pi^-})^2$ is
the square of the center of mass energy of the charged
pion pair. Performing the change
of basis (\ref{change of basis}), one finds that $b$ is given by
$-\epsilon/F^2$, where $\epsilon$ is one of the mixing angles introduced
above. The amplitude may thus be written as
\bdm A=-\epsilon\,\frac{1}{F^2}(s -\sA)\fs\edm
The result is of the same structure as the current algebra prediction
\cite{Cronin},
\bdm A=-\epsilon_0\,\frac{1}{F^2}(s -\mbox{$\frac{4}{3}$}M_\pi^2)\fs\edm
As a consequence of the interference with the $\eta'$, the
quark mass ratio $\epsilon_0$ is replaced by the mixing angle $\epsilon$ given
in eq.(\ref{epsilon}). There is a corresponding change also in the value of the
constant term, $\frac{4}{3}M_\pi^2\!\rightarrow
\sA$, but this term is inessential,
for the following reason. In the limit $m_u,m_d\!\rightarrow\!0$,
the amplitude contains two Adler zeros, one at
$p_{\pi^+}\!=\!0$,
the other at $p_{\pi^-}\!=\!0$. For the above expression to have this
property, the constant $\sA$ must tend to zero if $m_u,m_d$ are turned off.
Hence an explicit evaluation would yield a result for $\sA$ proportional to
$M_\pi^2$: The
interference with the $\eta'$ merely generates an SU(3) correction in the value
$\frac{4}{3}$ of the coefficient. Since contributions of order $M_\pi^2$ amount
to small corrections, I will drop these in the following.

Using the observed values of
$M_\pi^2,M_K^2,M_{\eta'}^2-M_\eta^2$ as an input, the relation (\ref{mixing a})
yields $\theta_{\eta\eta'}\simeq-22^\circ $, in reasonable
agreement with what is found phenomenologically \cite{Phenomenology}.
Inserting this number, formula (\ref{epsilon}) gives
$\epsilon\simeq2\,\epsilon_0$. So, the net result of the above
calculation is that mixing
with the $\eta^\prime$ increases the current algebra prediction for
the amplitude by a factor of 2.

I now compare this finding with the one loop result of chiral perturbation
theory \cite{GL eta}. This calculation accounts for all effects of first
nonleading order, in particular also for those due to $\eta\eta'$ mixing.
It is based
on SU(3)$_\R\times$SU(3)$_\L$ and hence only involves the degrees of freedom of
the pseudoscalar octet. In this framework, the $\eta^\prime$ only manifests
itself indirectly, through its contributions to the effective coupling
constants, like all other
states which remain massive in the chiral limit, e.g. the $\rho$.

Normalizing the amplitude with the kaon mass
difference (e.m. self energies removed) and with the pion matrix element of
the axial current,\footnote{Since the expression accounts for the
corrections
of order $m$, one needs to distinguish between the constant $F$
in the effective Lagrangian and the observed decay
constants $F_\pi,F_K$, which differ from $F$ through contributions of
order $m$.}
the result is of the form \cite{GL SU(3)}
\be\label{one loop}   A=-\frac{(M_{K^0}^2-M_{K^+}^2)}
{{3\sqrt{3}\,F_\pi}^{\hspace{-0.2em}2}}\,M(s,t,u)\co\ee
where $M(s,t,u)$ is a lengthy expression, which contains contributions
generated by the final state interaction, as well as
symmetry breaking terms involving the effective coupling constants
$L_5,L_7,L_8$. The $\eta^\prime$ hides in the coupling
constant $L_7$,
which also occurs if the mass of the $\eta$ is calculated within the same
framework. The explicit expression for the function $M(s,t,u)$
contains this constant through a correction term which is
proportional to the deviation from the Gell-Mann-Okubo formula and is denoted
by
\bdm \DGMO\equiv\frac{4 M_K^2-3M_\eta^2-M_\pi^2}{M_\eta^2-M_\pi^2}\fs\edm
Dropping all other terms and
disregarding contributions of order $M_\pi^2$, the one loop result reduces to
\bdm
M(s,t,u)=\frac{3\,s}{M_\eta^2\!-\!M_\pi^2}(1+\mbox{$\frac{2}{3}$}\DGMO)
+\ldots=\frac{9\,s}{4(M_K^2\!-\!M_\pi^2)}(1+\DGMO)+\ldots\co
\nonumber\edm where I have used the identity
$(M_\eta^2\!-\!M_\pi^2)(1+\frac{1}{3}\DGMO)=\frac{4}{3}
(M_K^2\!-\!M_\pi^2)$. The current algebra mass formulae quoted above show
that, at
leading order of the chiral perturbation series, the ratio
$(M_{K^0}^2-M_{K^+}^2)/(M_K^2-M_\pi^2)$ is given by
$(m_d-m_u)/(m_s-\hat{m})\!=\!4\epsilon_0/\sqrt{3}$.
Since the corresponding first order corrections do not involve the coupling
constant $L_7$, they are irrelevant in the present context and the same
also holds for the difference between $F_\pi$ and $F$.

With these simplifications, eq.(\ref{one loop}) reduces to
$A=-(\epsilon_0/F^2)\, s\, (1+\DGMO)$: Up to small corrections of order
$M_\pi^2$, the contribution from the symmetry breaking terms
amounts to an overall renormalization of the amplitude,
$\epsilon_0\!\rightarrow\!\epsilon_0\,(1+\DGMO)$.
The experimental value of the
deviation from the Gell-Mann-Okubo formula, $\DGMO\!=\!0.22$, shows that
the modification is of reasonable size, confirming
the general rule of thumb, according to
which first order SU(3) breaking
effects are typically of order $25\%$. The second order
contributions, which the one loop formula neglects, are expected to be of the
order of the square of this. Clearly, the outcome of the calculation
described earlier is in flat contradiction with chiral perturbation theory.

To identify the origin of the disagreement, I return to the earlier
calculation and express the angular
factor occurring in eq.(\ref{epsilon}) in terms of the masses of the particles.
Solving the relation (\ref{mixing b}) for
$\mbox{tg}\,\theta_{\eta\eta'}$ and inserting the result, the
angular factor becomes
\bdm \frac{\cos\mixingangle
-\sin \mixingangle\,\sqrt{2}}{\cos\mixingangle+
\sin \mixingangle/\sqrt{2}}
=\frac{2(2M_K^2-M_\eta^2-M_\pi^2)}{M_\eta^2-M_\pi^2}\equiv 1+\DGMO\fs\edm
This is remarkable, because it shows that the expression for
the mixing angle $\epsilon$
may equally well be written as
\be\label{angular factor} \epsilon=
 \epsilon_0\,
\,\{1+\Delta_{\mbox{\tiny GMO}}\}\cos \mixingangle \fs\ee
In this form, the result of the direct calculation differs from
the corresponding term in the one loop prediction of chiral perturbation theory
only by a factor of $\cos\mixingangle\!\simeq\!0.93$, which
represents a correction of order $(m_s-\hat{m})^2$ and is beyond the accuracy
of the one loop result. I conclude that
the two calculations are consistent with one another.
In particular, it is incorrect to amalgamate
the two by multiplying the
one loop formula with the enhancement factor occurring in eq.(\ref{epsilon}).

The above expression for the angular factor shows that the result of the
direct calculation is subject to an uncertainty comparable to the effect
itself:
Depending on whether one uses eq.(\ref{mixing a}) or eq.(\ref{mixing b}), i.e.
takes the observed values of $M_{\eta'}^2-M_\eta^2$ or $M_\eta^2$ as an
input, the calculation yields $\epsilon\simeq
2\,\epsilon_0$ or $\epsilon\simeq 1.2\,\epsilon_0$, respectively.
The problem arises from
the fact that the mass formula (\ref{mixing b}) is not in good
agreement with observation.
If the second term is omitted, the relation reduces to the Gell-Mann-Okubo
formula, which predicts $M_\eta\!\simeq\! 566\,\mbox{MeV}$, slightly
larger than what is observed. The second term indeed lowers the result,
but the shift is much too large: While the Gell-Mann-Okubo prediction for
$M_\eta^2$ only differs from the experimental value by 7 \%, the repulsion
generated by mixing now yields a number which is too low by about 20 \%.
As pointed out in ref.\cite{GL SU(3)}, early
determinations of the mixing angle failed for precisely this reason:
These were based on the assumption that
the observed deviation from the Gell-Mann-Okubo formula is exclusively due to
$\eta\eta'$
mixing and thus underestimated the magnitude of $\mixingangle$ by about a
factor of two.

The above discrepancies do not indicate that the expansion of the effective
Lagrangian in powers of $1/N_c,p$ and $m$ fails. Deviations of this order
of magnitude are to be expected within a framework which only considers the
leading term of the expansion.
The effective Lagrangian in eq.(\ref{Leff})
also predicts that $F_K$ is equal to $F_\pi$, while, experimentally, the
two quantities differ by the factor 1.22. There is no reason
why in the case of the masses, the corrections generated by the higher order
contributions of the expansion should be smaller.

The observed mass pattern is
perfectly consistent
with the assumption that the terms neglected in eq.(\ref{Leff})
are small, but they definitely are different from zero.
The main point here is that the same terms necessarily also
affect the
amplitude of the transition $\eta\rightarrow 3\pi$. The
result of the chiral perturbation theory calculation
amounts to a low energy theorem: To order $p^4$,
the slope of the decay amplitude
involves the same combination of couplings which
determines the deviation from the Gell-Mann-Okubo formula.
While the effective Lagrangian in eq.(\ref{Leff}) only accounts for the
coupling constant $L_7$, which is related to $\eta\eta'$ mixing, the one loop
result receives significant contributions also from $L_5$ and $L_8$.
In the framework
of pole models \cite{Ecker Gasser Pich de Rafael}, these couplings are
dominated by the exchange of scalar particles.
Indeed, the particle data table shows that the mass of the lightest scalars
is comparable with $M_{\eta'}$.
These particles do not undergo
mixing with pseudoscalar one-particle states, but with the ground state as well
as with two-particle states.
The corresponding effect in the square of the
pseudoscalar masses is also of order $m^2$ and is of opposite sign. It is
suppressed by a relatively large
energy denominator, because the masses of the scalars are large compared to
those of the pseudoscalar octet, but the energy denominator is essentially
the same as the one which suppresses the shift generated by
$\eta\eta^\prime$ mixing, so that the effects are of the same order of
magnitude.

This explains why the direct calculation does not yield a decent estimate for
$\epsilon$, unless the result is written in the form (\ref{angular factor}),
where it differs from the chiral perturbation theory result only by a factor of
$\cos \mixingangle$. As noted above, this factor represents a second order
correction of typical size. I conclude that there is no
indication for the symmetry breaking effects of higher order to be
unusually large -- the factor $\cos\mixingangle$ may be taken as an
estimate for the uncertainties in the decay amplitude due to these.

The current algebra prediction for the decay $\eta\!\rightarrow\! 3\pi$ also
receives corrections from a quite different source: final state interactions.
These are generated predominantly by two-particle branch cuts and are
responsible for the bulk of the one loop corrections.
The corresponding higher order contributions may be
worked out from unitarity, using dispersion relations and relying on chiral
perturbation theory only to determine the subtraction constants
\cite{Kambor Wiesendanger Wyler,AL}. Viewed in this
perspective, the above discussion implies that the one loop predictions for the
subtraction constants are trustworthy: These account for all
symmetry breaking effects of first nonleading order, in particular for those
due to $\eta\eta'$ mixing, and there is no indication for large corrections
from higher orders.

Note that these statements need not hold for radiative transitions,
which may well be distorted by the pole due to $\eta^\prime$-exchange
\cite{Akhoury Leurer Pich,Diakonov}. The $3\pi$ channel
is special, because the transition amplitude is determined by the effective
Lagrangian of the strong
interaction --- this is why it is firmly tied to the
mass spectrum of the pseudoscalars.

\end{document}